\documentclass[12pt]{article}
\usepackage{fullpage}
\usepackage[utf8]{inputenc} 
\usepackage[T1]{fontenc}
\usepackage{amsfonts}      
\usepackage{graphicx}
\usepackage{pdfpages}
\usepackage{amsmath}
\usepackage{amsthm}
\usepackage{amssymb}
\usepackage{natbib}
\usepackage{tikz}
\usepackage{algorithm}
\usepackage{algorithmic}
\usepackage{enumitem}

\newcommand{\up}{\text{Up}}
\newcommand{\dw}{\text{Dw}}

\newcommand\crules[3][black]{\textcolor{#1}{\rule{#2}{#3}}}
\definecolor{fig.labeled.data.grey}{RGB}{229,229,229}
\definecolor{fig.labeled.data.blue.start}{RGB}{99,184,255}
\definecolor{fig.labeled.data.blue.end}{RGB}{65,105,225}
\definecolor{fig.labeled.data.blue.correct}{RGB}{34,139,34}
\definecolor{fig.labeled.data.blue.FN}{RGB}{255,165,0}
\definecolor{fig.labeled.data.blue.FP}{RGB}{238,0,0}
\definecolor{max_accuracy}{RGB}{179, 235, 168}
\definecolor{royalblue}{RGB}{65, 105, 225}
\definecolor{poisson}{RGB}{230, 159, 0}
\definecolor{negbin}{RGB}{156, 206, 55}
\definecolor{gaussian}{RGB}{86, 180, 233}

\title{Increased peak detection accuracy in over-dispersed ChIP-seq data with supervised segmentation models}

\author{
   Arnaud Liehrmann,
  \texttt{arnaud.lieh@gmail.com} \footnote{Université Paris‐Saclay, CNRS, INRAE, Univ Evry, Institut des Sciences des Plantes de Paris-Saclay (IPS2), Orsay, 91405, France.} \footnote{Université Paris-Saclay, CNRS, Univ Evry, Laboratoire de Mathématiques et Modélisation d'Evry, Evry, 91037, France.}\\
  Guillem Rigaill, 
  \texttt{guillem.rigaill@inrae.fr} \footnotemark[1] \footnotemark[2]\\
  Toby Dylan Hocking, 
  \texttt{toby.hocking@nau.edu} \footnote{Northern Arizona University, School of Informatics, Computing, and Cyber Systems 1295 S. Knoles Dr., Building 90, Room 210, Flagstaff, AZ, 86011, USA.}
}
\date{}

\begin{document}

\maketitle
\vspace*{0.7cm}
\begin{abstract}
 \noindent\textbf{Motivation:}$\quad$ Histone modification constitutes a basic mechanism for the genetic regulation of gene expression. In early 2000s, a powerful technique has emerged that couples chromatin immunoprecipitation with high-throughput sequencing (ChIP-seq). This technique provides a direct survey of the DNA regions associated to these modifications. In order to realize the full potential of this technique, increasingly sophisticated statistical algorithms have been developed or adapted to analyze the massive amount of data it generates. Many of these algorithms were built around natural assumptions such as the Poisson one to model the noise in the count data. In this work we start from these natural assumptions and show that it is possible to improve upon them.

\noindent\textbf{Results:}$\quad$ The results of our comparisons on seven reference datasets of histone modifications (H3K36me3 \& H3K4me3) suggest that natural assumptions are not always realistic under application conditions. We show that the unconstrained multiple changepoint detection model with alternative noise assumptions and a suitable setup reduces the over-dispersion exhibited by count data and turns out to detect peaks more accurately than algorithms which rely on these natural assumptions.
\end{abstract}
\textbf{Keywords:}$\quad$ChIP-seq; histone modifications; over-dispersion; peak calling; multiple changepoint detection; likelihood inference; supervised learning
\newpage
\section{Introduction}

\paragraph{Context.} Chromatin immunoprecipitation followed by high-throughput sequencing (ChIP-seq) is amongst the most widely used methods in molecular biology \citep{mari18}. This method aims to identify transcription factor bindings sites \citep{val08, sch10} or post-translational histone modifications \citep{you11, zha16}, referred to as histone marks, underlying regulatory elements. Consequently, this method is essential to deepen our understanding of transcriptional regulation. The ChIP-seq assay yields a set of DNA sequence reads which are aligned to a reference genome and then counted at each genomic position. This results in a series $Y = (y_1,\dots,y_n)$ of $n$ non-negative integer count data $(y_i \in \mathbb{Z}_+)$, hereafter called coverage profile, ordered along a chromosome. The binding sites or histone marks of interest appear as regions with high read density referred to as peaks in the coverage profile. 

Since there is a biological interest in detecting these peaks, several methods, hereafter called peak callers ($c$), have been developed / adapted and used to filter out background noise and accurately identify the peak locations in the coverage profile. They take a coverage profile of length $n$ and classify each base from it as a part of the background noise (0) or peak (1), i.e. $c:Y\to\{0,1\}^n$. Among these peak callers we can mention MACS \citep{zhan08} and HMCan \citep{asho13}, two heuristics which are computationally fast but typically accurate only for a specific pattern, i.e. respectively sharp and broad peaks \citep{toby16}. More recently, it has been proposed to solve the peak detection problem using either optimal constrained or unconstrained multiple changepoint detection methods \citep{toby15, toby20}. The constraints ensure that the segmentation model can be interpreted in terms of peaks and background noise which is a practitioner's request. The unconstrained one doesn't have an output segmentation with a straightforward interpretation in terms of peaks and needs to be followed by an ad-hoc post-processing rule to infer the start and end of peaks (see Section \ref{alieh:PDPA.rules.explain}). For each of these methods, there are one or more tuning parameters that need to be set before solving the peak detection problem and that may affect the results accuracy \citep{toby16, toby20}. 

In a supervised learning approach, \citet{toby16} introduced seven labeled histone mark datasets that are composed of samples from two different ChIP-seq experiments directed at histone modifications H3K36me3 and H3K36me3 (see Section \ref{dataset}). In a recent study, after training different peak callers using these datasets, \citet{toby20} compared them and showed that the constrained segmentation model with coverage data following a Poisson distribution outperforms standard bioinformatics heuristics and the unconstrained segmentation model on these datasets.
\paragraph{Modeling question.} From a modeling perspective the constrained segmentation model and the Poisson noise are certainly the most natural assumptions to detect peaks in count data. However, it is not clear that they are realistic:
\begin{itemize}
    \item By looking at the shapes of the peaks in coverage profiles (see for instance in figure \ref{alieh:presentation.models}), we can see that the background noise and the top of them are sometimes separated by one or more subtle changes. In contrast to the constrained segmentation model, the unconstrained one should be able to capture these subtle changes. One major issue is that the output segmentation of the unconstrained model does not have a straightforward interpretation in terms of peaks.
    \item Parametric models such as the negative binomial \citep{robi10, love14} or the Gaussian, following a proper transformation of the count data for the latter \citep{ansc48, law14}, are preferred over the Poisson one for the analysis of many high-througput sequencing datasets. Indeed, count data often exhibit more variability than the Poisson model expects which changes the interpretation of the model and makes it difficult to estimate its parameters. These alternative parametric models are well known to reduce this phenomenon, also called over-dispersion.
\end{itemize}
In this work we try to start from these natural assumptions and show that it is possible to improve upon them.  
\paragraph{Contribution.} 
\begin{enumerate}
    \item We show that the distribution of counts from H3K36me3 and H3K4me3 datasets exhibits over-dispersion which invalidates the Poisson assumption. The two alternative noise models we propose (negative binomial with constant dispersion parameter \& Gaussian after Anscombe transformation) effectively reduce the over-dispersion  on these datasets (see Figure \ref{alieh:empirical.variance.vs.theoretical.variance}).
    
    \item We propose a new and rather natural post-processing rule to predict the start and end of peaks in an estimated unconstrained segmentation (see Section \ref{alieh:PDPA.rules.explain}). Indeed, in the unconstrained segmentation we can observe several up (respectively down) changes and it is not obvious which one should be considered as the start or end of the peak. We show that this new post-processing rule improves the accuracy of this segmentation model in both H3K36me3 and H3K4me3 datasets compared to the same model with previous rules described by \citet{toby20} (see Figure \ref{alieh:res.rules}).

    \item \citet{toby18} described a procedure to extract all optimal constrained segmentations for a range of peaks. It is an essential internal step in the supervised approach for learning the penalty parameter of segmentation models. In this work we generalize this procedure so that it works with the unconstrained segmentation model and the post-processing rule mentioned in the previous point (see Algorithm \ref{alieh:algo1}).
    
    \item We describe a method to learn jointly both the penalty and dispersion parameters of segmentation models with a negative binomial noise (see Section \ref{supervised.learning}). We then compare the accuracy of unconstrained and contrained segmention models with different noise distributions on the labeled H3K36me3 and H3K4me3 datasets (see Figure \ref{alieh:res.comp}).
\end{enumerate}

\section{Segmentation models for ChIP-seq data}
\begin{figure}
\centering
\includegraphics[scale=0.4]{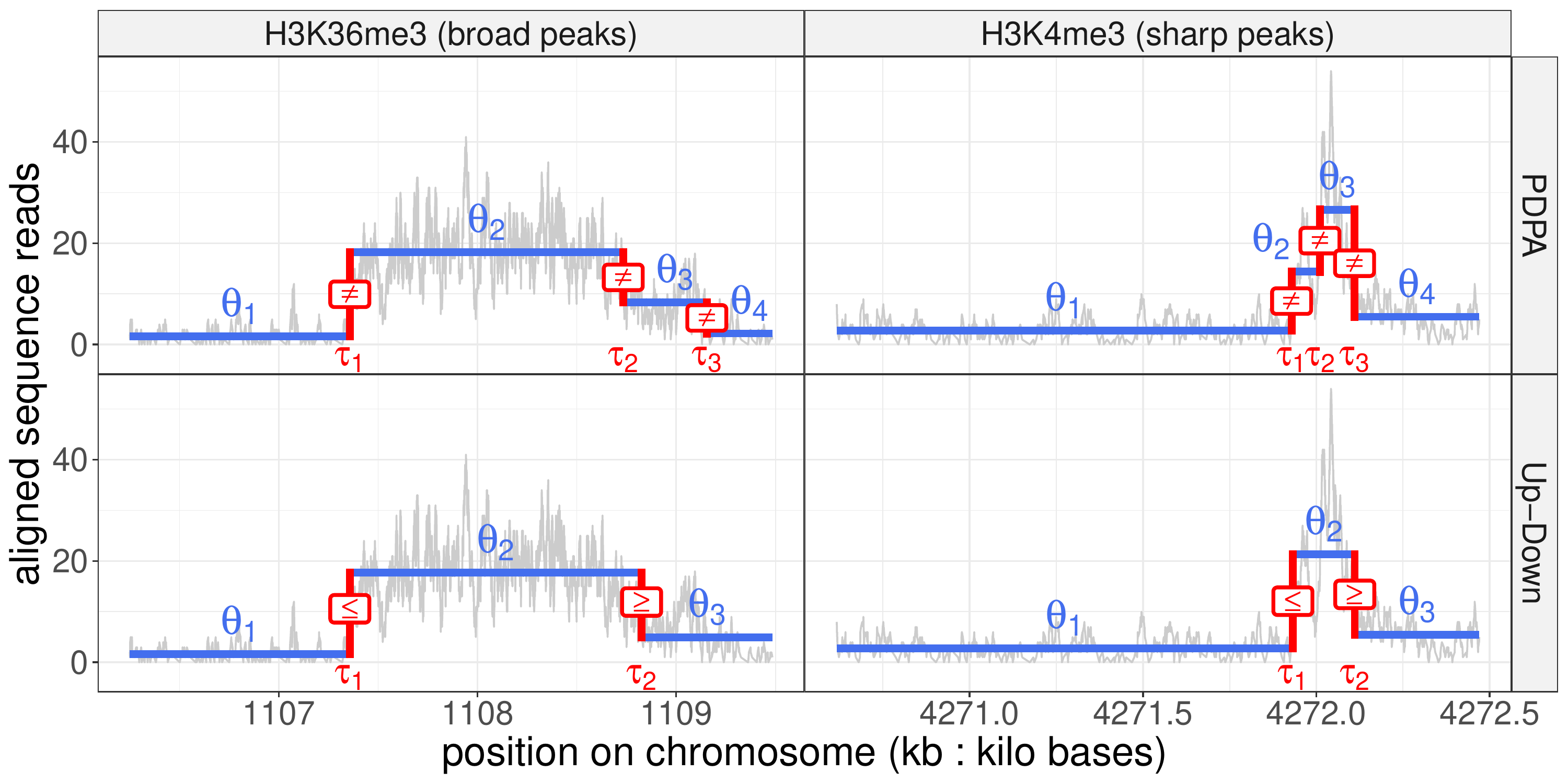}
\caption{Examples of ChIP-seq coverage profiles from the histone mark H3K36me3 and H3K4me3 datasets. (\textbf{Top}) In blue (\crules[royalblue]{0.25cm}{0.25cm}), a piecewise constant function affected by three unconstrained abrupt changes shown in red (\crules[red]{0.25cm}{0.25cm}). (\textbf{Bottom}) In blue (\crules[royalblue]{0.25cm}{0.25cm}), a piecewise constant function affected by two constrained abrupt changes shown in red (\crules[red]{0.25cm}{0.25cm}).}
\label{alieh:presentation.models}
\end{figure}
\paragraph{Unconstrained segmentation model.} The observed data $(y_1,\dots,y_n)$ are supposed to be a realization of an independent random process $(Y_1,\dots,Y_n)$. This process is drawn from a probability distribution $\mathcal{F}$ which depends on two parameters: $\theta$ is assumed to be affected by $K-1$ abrupt changes called changepoints and $\phi$ is constant. We denote $\tau_k$ the location of the $k^{th}$ changepoint with $k=\{1,\dots,K-1\}$. By convention we introduce the fixed indices $\tau_0=0$ and $\tau_K=n$. The $k^{th}$ segment is formed by the observations $(y_{\tau_{k-1}+1} ,\dots, y_{\tau_k})$. $\theta_k$ stands for the parameter of the $k^{th}$ segment (see Figure \ref{alieh:presentation.models}). Formally the unconstrained segmentation model \citep{cley14}, hereafter called PDPA, can be written as follows:
\begin{equation} \label{alieh:unconstrained.model}
    \forall i\;|\quad \tau_{k-1}+1 \leq i \leq \tau_{k},\quad Y_{i} \sim \mathcal{F}(\theta_k, \phi)\quad.
\end{equation}
\paragraph{Constrained segmentation model.} In order to have a segmentation model with a straightforward interpretation in terms of peaks, we add inequality constraints to the successive segment specific parameters $(\theta_{1},\dots,\theta_{K})$ so that non-decreasing changes in these parameters are always followed by non-increasing changes. Therefore, we formally assume the following constrained segmentation model \citep{toby15}, hereafter called Up-Down:
\begin{align}
    \forall i\; |\quad \tau_{k-1}+1 \leq i \leq \tau_{k},\quad Y_{i} \sim \mathcal{F}(\theta_k, \phi)\quad \\\text{subject to} 
    \begin{cases}
    \theta_{k-1} \leq \theta_{k} \quad \forall k \in \{2,4,\dots\} \\
    \theta_{k-1} \geq \theta_{k} \quad \forall k \in \{3,5,\dots\}
\end{cases}. \nonumber
\end{align}
\paragraph{Probability distributions.} In the case of the Poisson distribution we have $\mathcal{F}(\theta_k, \phi) = \mathcal{P}(\Lambda_k, \phi=\emptyset)$ where $\Lambda_k$ stands for the mean and the variance of the $k^{th}$ segment. In the case of the Gaussian distribution we have $\mathcal{F}(\theta_k, \phi) = \mathcal{N}(\mu_k,\sigma^2)$ where $\mu_k$ is the mean of the $k^{th}$ segment and $\sigma^2$ is the variance assumed constant across the segments. Also in this case, the non-negative integer count data have been transformed in real values $(\mathbb{Z}_+ \to \mathbb{R}_+)$ through an Anscombe transformation $(Y = \sqrt{Y+\frac{3}{8}})$ which is a useful variance-stabilizing transformation for over-dispersed count data following a Poisson distribution \citep{ansc48}. In the case of the negative binomial distribution we have $\mathcal{F}(\theta_k, \phi) = \mathcal{NB}(p_k,\phi)$ which is the natural parametrization from the exponential family where $\phi$ is the dispersion parameter that needs to be learned on the data and $p_k$ is the probability of success, $0\leq p_k \leq 1$, and is the inverse of the mean of the $k^{th}$ segment ($p_k = \frac{\phi}{\phi + \mu_k}$). For the Up-Down model, the inequality constraints are applied on $1-p_k$ to keep their interpretation consistent with the piecewise constant mean model we consider.
 \paragraph{Optimization problems.} \label{optimization.problem} In both unconstrained and constrained optimal multiple changepoint detection problems, the goal is to estimate the changepoint locations $(\tau_1,\dots,\tau_{K-1})$ and the parameters $(\theta_1,\dots,\theta_{K})$ both resulting from the segmentation. \citet{rung20} introduced \textit{gfpop}, an algorithm that solves both problems using penalized maximum likelihood inference. It implements several loss functions including the Gaussian, Poisson and negative binomial that allowed us to compare different noise models for the count data. The number of changepoints in a coverage profile being unknown, \textit{gfpop} takes a non-negative penalty $\lambda \in \mathbb{R}_+$ parameter that controls the complexity of the output segmentation. Larger penalty $\lambda$ values result in models with fewer changepoints. The extreme penalty values are $\lambda = 0$ which yields $n-1$ changepoints, and $\lambda = \infty$ which yields $0$ changepoint. The time complexity of \textit{gfpop} is empirically $\mathcal{O}(Vn\log(n))$. Intuitively, $V$ stands for the number states you will need to encode your priors about the form of the output segmentation, e.g. with the Up-Down model at each time the signal can be a part of the background noise (Down) or a peak (Up). Consequently, the empirical time complexity of \textit{gfpop} with the Up-Down model is $\mathcal{O}(2n\log(n))$ while with the PDPA model it is $\mathcal{O}(n\log(n))$.

\section{Rules for inferring the start and end of peaks with the unconstrained segmentation model (PDPA)}\label{alieh:PDPA.rules.explain}
\begin{figure}
\centerline{\scalebox{0.5}{\input{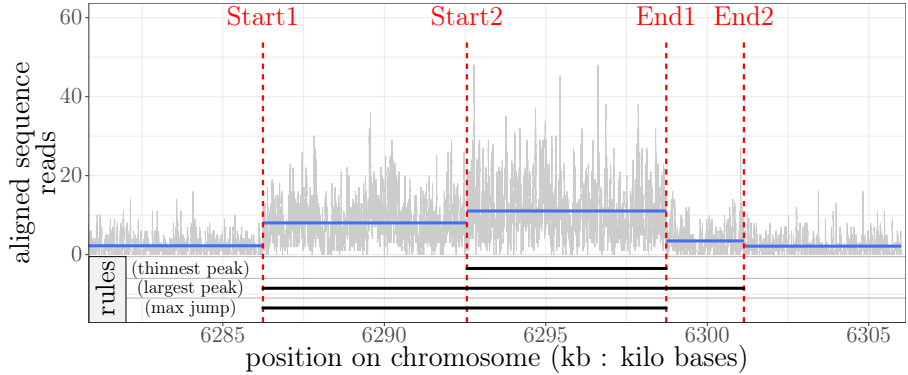}}}
\caption{(\textbf{Top}) Segmentation of a coverage profile containing one peak using the PDPA model. The location of the changepoints on the chromosome are shown by red dotted lines (\crules[red]{0.25cm}{0.25cm}). The mean of the segments are shown in blue (\crules[royalblue]{0.25cm}{0.25cm}). According to this segmentation there are two alternative starts and two alternative ends of the peak, i.e. four alternative variants of the same peak formed by the regions: [Start1:End1], [Start1:End2], [Start2:End1] and [Start2:End2]. (\textbf{Bottom}) Three different rules are proposed to interpret the segmentation as peaks. \textit{Thinnest peak}: the resulting peak is defined by the region [Start2:End1]. \textit{Largest peak}: the resulting peak is defined by the region [Start1:End2]. \textit{Max jump}: the resulting peak is defined by the region [Start1:End1].}
\label{alieh:PDPA.rules.explain.fig}
\end{figure}
As mentioned before, one of the main motivation of the Up-Down model is that it can be interpreted in terms of peaks which is a practitioner's request. In the case of the PDPA model, the output segmentation may results in successive non-decreasing changes ($\up^*$), e.g. in figure \ref{alieh:PDPA.rules.explain.fig}: $\up^*=\{\text{Start1}, \text{Start2}\}$, and successive non-increasing changes ($\dw^*$), e.g. in figure \ref{alieh:PDPA.rules.explain.fig}: $\dw^*=\{\text{End1}, \text{End2}\}$, in the signal. Thus, it is necessary to specify a post-processing rule to select the start and end of peaks among the returned changepoints in respectively each $\up^*$ and $\dw^*$. This results in $|\up^*| \times |\dw^*|$ alternatives of the same peak. 
\paragraph{Rules.} We propose three different rules to select the start and end of peaks (see Figure \ref{alieh:PDPA.rules.explain.fig}): 
\begin{itemize}[leftmargin=0.5cm]
\item\textit{thinnest peak}: we select the last up change in $\up^*$ and the first down change in $\dw^*$ ; 
\item\textit{largest peak rule}: we select the first up change in $\up^*$ and the last down change in $\dw^*$ ; 
\item\textit{max jump}: we select the up and down change with the largest mean-difference in $\up^*$ and $\dw^*$.
\end{itemize}
\citet{toby20} introduced similar rules to the \textit{thinnest peak} and \textit{largest peak}.

\section{Labeled data for supervised learning peak detection}
\paragraph{Tuning parameters.} For each peak callers there are one or more tuning parameters that need to be set before solving the peak detection problem and that may greatly affect the result accuracy. For segmentation methods this parameter is the penalty $\lambda$ which controls the number of peaks in the resulting segmentation, while for heuristics, such as MACS or HMCan, they use a threshold parameter whose value allows to only consider the top $p$ peaks given their significance. Moreover, if we want to model the over-dipersion phenomenon observed in the count data using a negative binomial probability distribution, this is done at the cost of another parameter $(\phi)$ that we need to set as well. Its value may also affect the number of peaks in the resulting segmentation. In theory, if the correct noise model was known, it would be possible to use statistical arguments to choose the parameter to use. However, in practice the correct noise model is complex and unknown. There are many factors that influence the signal and noise patterns in real ChIP-seq data, e.g. experimental protocols, sequencing machines, alignment software, auto-correlation. These factors results in poor accuracy for the detection of peaks \citep{toby16}. Therefore, we will consider the supervised peak detection problem in which the value of tuning parameters can be learned using manually determined labels that indicate a presence or absence of peaks.
\begin{figure}
\centerline{\scalebox{0.47}{\input{figure_labeled_data}}}
\caption{(\textbf{Top}) Example of a ChIP-seq coverage profile annotated by an expert biologist. The labels represented by colored rectangles indicate the absence (\crules [fig.labeled.data.grey]{0.25cm}{0.25cm}) or presence of a peak, here characterized by its start (\crules [fig.labeled.data.blue.start]{0.25cm}{0.25cm}) and its end (\crules [fig.labeled.data.blue.end]{0.25cm}{0.25cm}). (\textbf{Bottom}) The model with 1 peak in its output segmentation has an associated error of 2 ($2\times$ False Negative \crules [fig.labeled.data.blue.FN]{0.25cm}{0.25cm}). The model with 3 peaks has an associated error of 1 ($1\times$ False Positive \crules [fig.labeled.data.blue.FP]{0.25cm}{0.25cm}). The model with 2 peaks is a good model for which all the labels optimized on this coverage profile are correct  (\crules[fig.labeled.data.blue.correct]{0.25cm}{0.25cm}).
}
\label{alieh:labeled.data}
\end{figure}
\paragraph{Benchmark datasets.}\label{dataset} Introduced by \citet{toby16}, these seven labeled histone mark datasets are composed of samples from two different ChIP-seq experiments directed at modifications found on the histone 3 N-terminal tails. The first experiment is directed at histone H3 lysine 4 tri-methylation (H3K4me3), a modification localized in promoters. The second one is directed at histone H3 lysine 36 tri-methylation (H3K36me3), a modification localized in transcribed regions. Both these modifications are both often involved in the regulation of gene expression \citep{Sims03}. The histone modifications H3K4me3 and H3K36me3 are respectively characterized by sharp and broad peak patterns in coverage profiles. Expert biologists, with visual inspection, have annotated some regions by indicating the presence or absence of peaks. Then, they grouped the labels to form 2752 distinct labeled coverage profiles. 

\paragraph{Definition of labeled coverage profiles and errors.} In the context of supervised peak detection each labeled coverage profile of size $n$, denoted $w\in\mathbb{Z}_+^n$, is a problem. Formally we have a set of $M$ problems $(w_1,\dots,w_M)$ where $M=2752$. Each problem $w_m$ is associated with a set of $N$ labels $H_m = \{(s_1,e_1, h_1)\, \dots, (s_{N},e_{N}, h_{N})\}$ where $s$ is the start genomic location of the label, $e$ is the end genomic location of the label and $h$ is the type of the label. There are four types of labels that allow some flexibility in the annotation (see Figure \ref{alieh:labeled.data}): 
\begin{itemize}
\item \textit{noPeaks} label stands for a region that contains only background noise with no peak. If any peak is predicted in this region, the label counts as a false positive ; 
\item \textit{peaks} label means there is at least one overlapping peak in that region. Hence, one or more peaks in that region is acceptable. If there is not at least one overlapping peak predicted in this region, the label counts as a false negative ; 
\item \textit{peakStart} and \textit{peakEnd} labels stand for regions which should contain exactly one peak start or end. If more than one peak start / end is predicted in this region, the label counts as a false positive. Conversely, if less than one peak start / end is predicted in this region, the label counts as a false negative.
\end{itemize}
The set of labels $H_m$ is used to quantify the error $E_m$, i.e. the total number of incorrectly predicted labels (false positive $+$ false negative) in the coverage profile $w_m$ given the set of peaks returned by a peak caller. 

\section{Supervised algorithms for learning tuning parameters of negative binomial segmentation models} \label{supervised.learning}
\paragraph{Objective function.} The error function for a given problem $w_m$, denoted $E_m:\mathbb{R}_+^2\to\mathbb{Z_+}$, is a mapping from the tuning parameters $(\phi, \lambda$) of negative binomial segmentation models to the number of incorrectly predicted labels in the resulting optimal segmentation. With the supervised peak detection approach, the goal is to provide predictions of $\phi$ and $\lambda$ that minimize $E_m(\phi,\lambda)$. The exact computation of the 2-dimensional defined $E_m(\phi,\lambda)$ is intractable with respect to $\phi$. Thus, we computed it over 16 $\phi$ values evenly placed on the log scale between 1 and 10000, $\Phi = (\phi_1=1,\dots,\phi_{16}=10000)$. Initial results suggest that this grid of values is a good set of candidates to test in order to calibrate the dispersion parameter $\phi$ (see Supplementary Figure 1). The exact computation of the error rate as a function of $\lambda$ ($\phi$ remains constant), a piecewise constant function, requires to retrieve all optimal segmentations up to 9 peaks. This way, on the advice of the biologists who annotated the benchmark datasets, we ensure that for each problem there is a segmentation with at least one false positive / false negative label. A procedure that retrieves one optimal segmentation for each targeted number of peaks $P^*$ has already been described by \citet{toby18}. It can be used with the Up-Down model for which there is at most one optimal segmentation that results in $P^*$ peaks but not with the PDPA model for which there can be several ones.  Indeed, the constraints in the Up-Down model require it to add, if the associated cost is optimal, 2 changepoints that lead to the formation of a new peak. With the PDPA model, adding a changepoint can either refine an already existing peak or, in combination with another changepoint, form a new peak. More generally there is a need of an algorithm that takes as input any penalized changepoint detection solver $\mathcal{S}$ with a penalty $\lambda$ constant along the changepoints and outputs all optimal segmentations between two peak bounds denoted $\underline{P}$ and $\overline{P}$. 
 We present \textit{CROCS} \includegraphics[scale=0.05]{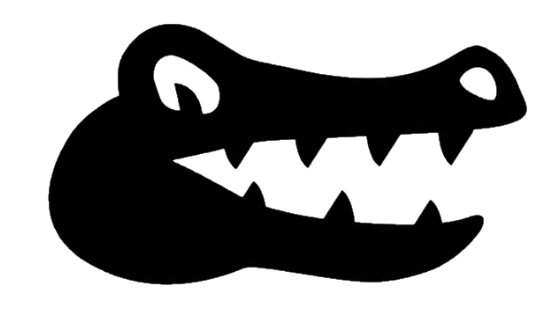} (Changepoints for a Range of ComplexitieS), an algorithm that meets this need. 
\paragraph{Discussion of pseudocode: \textit{CROCS} (Algorithm \ref{alieh:algo1}).}\textbf{(i)} The algorithm begins by calling \textit{SequentialSearch} \hyperlink{sequentialsearch}{[described underneath]} to search two penalty bounds $\overline{\lambda}$ (\textit{line 6}) and $\underline{\lambda}$ (\textit{line 5}) that result in a segmentation with respectively $\underline{P}-1$ (\textit{line 3}) and $\overline{P}+1$ (\textit{line 4}) peaks. Indeed, using \textit{gfpop} with the Up-Down model as solver $\mathcal{S}$, the number peaks in the resulting optimal segmentations is a non-increasing function of $\lambda$. This propriety guarantees that with the previous penalty bounds we can reach every optimal model from $\underline{P}$ to $\overline{P}$ peaks. For other unconstrained segmentation models, such as PDPA with the \textit{max jump} post-processing rule, we suspect it also should be true in the vast majority of cases. \textbf{(ii)} Then, the algorithm calls \textit{CROPS} \hyperlink{CROPS}{[described underneath]} (\textit{line 7}) to retrieve all the optimal segmentations between these two penalty bounds. \textbf{(iii)} Finally, a simple post-processing step (not shown in the algorithm) allows to remove segmentations with $P-1$ and $P+1$ peaks. The time complexity of the \textit{CROCS} \includegraphics[scale=0.05]{figure_crocs.png} algorithm is bounded by the time complexity of the \textit{CROPS} procedure, i.e. $\mathcal{O}(\mathcal{O}(\mathcal{S})(K_{\underline{\lambda}} - K_{\overline{\lambda}}))$, where $K_{\overline{\lambda}}$ and $K_{\underline{\lambda}}$ are the number of segments in optimal segmentations associated to respectively $\overline{\lambda}$ and $\underline{\lambda}$. $\mathcal{O}(\mathcal{S})$ is the time complexity of the solver $\mathcal{S}$, e.g. empirically $\mathcal{O}(2n\log(n))$ for \textit{gfpop} with the Up-Down model (see Section \ref{optimization.problem}).
\begin{itemize}
    \item \textbf{\textit{SequentialSearch}} \hypertarget{sequentialsearch}{is a procedure described by \citet{toby18} that takes as input a problem $w_m$, a target number of peaks $P^*$ and outputs an optimal segmentation with $P^*$ peaks in addition to the penalty $\lambda$ for reaching it.} 
    \item \textbf{\textit{CROPS}} \hypertarget{CROPS}{is a procedure described by \citet{hayn14} that takes as input a problem $w_m$, as well as two penalty bounds $\underline{\lambda}$ \&  $\overline{\lambda}$ and outputs all the optimal segmentations between these two bounds.}
\end{itemize}
We slightly modified the original implementation of both \textit{SequentialSearch} and  \textit{CROPS} in such way that they can work with any penalized changepoint detection solver $\mathcal{S}$ provided by the user. The source code of \textit{CROCS} \includegraphics[scale=0.05]{figure_crocs.png} is available here: \url{https://github.com/aLiehrmann/chip_seq_segmentation_paper/tree/master/scripts/CROCS.r}

\begin{algorithm}
    \begin{algorithmic}[1]
\STATE \textbf{Input}: Data $w_m$, lower bound $\underline{P}$, upper bound $\overline{P}$, dispersion $\phi$, solver $\mathcal{S}$
\STATE \textbf{Output}: The details of optimal segmentations between $\underline{P}$ and $\overline{P}$ peaks
\STATE \textbf{if} $\underline{P}>0$: $\underline{P} \gets \underline{P} - 1$
\STATE $\overline{P} \gets \overline{P} + 1$
\STATE$\underline{\lambda} \gets \textit{SequentialSearch}(w_m,\overline{P},\mathcal{S})$ \hfill$\triangleright$ \citet{toby18}
\STATE $\overline{\lambda} \gets \textit{SequentialSearch}(w_m,\underline{P},\mathcal{S})$
\STATE \textbf{return} $\textit{CROPS}(w_m, \underline{\lambda}, \overline{\lambda}, \mathcal{S})$ \hfill$\triangleright$ \citet{hayn14}
\caption{\textit{CROCS} (Changepoints for a Range of ComplexitieS): extract all optimal segmentations between $\underline{P}$ and $\overline{P}$ using a changepoint penalized solver $\mathcal{S}$} \label{alieh:algo1}
\end{algorithmic}
\end{algorithm}
\vspace*{-.4cm}
\paragraph{Learning jointly $\phi$ and $\lambda$.} Once the error function $E_m(\phi\in\Phi,\lambda)$ is computed for each problem of the training set, a natural way to learn the dispersion and penalty parameters is to select the pair of values $(\phi\in\Phi,\lambda)$ that achieves the global minimum error. We denote these values $\phi^*$ and $\lambda^*$. Recall that $E_m(\phi\in\Phi,\lambda)$ is piecewise constant on $\lambda$. The sum of $E_m(\phi\in\Phi,\lambda)$ over all problems is still piecewise constant on $\lambda$. Therefore, $\phi^*$ and $\lambda^*$ can be easily retrieved using a sequential search. We refined the previous learning method, hereafter called \textit{constant $\lambda$}, by taking advantage of the piecewise constant propriety of  $E_m(\phi\in\Phi,\lambda)$. Indeed, the minimum error is not reached for a unique penalty value $\lambda^*$ but an interval denoted $I_{\lambda,m}$. After fixing $\phi^*$, we can use $I_{\lambda,m}$ computed for each problem of the training set in order to learn a function that predicts problem-specific $\lambda$ values. This function is a solution of the interval regression problem described by \citet{rigai13}. We denote this learning method \textit{linear $\lambda$}.

In the case of segmentation models with a Poisson or a Gaussian noise, the only tuning parameter that we need to learn is $\lambda$. Thus, the objective function becomes a 1-dimensional defined function denoted $E_m(\lambda)$. The methods we used to learn $\lambda$ are similar than those presented above (see \citet{toby20} for more details). 
\section{Empirical results}\label{results}
\paragraph{Cross-validation setup \& evaluation metric.} In the following section, for each model compared, a 10-fold or 4-fold \footnote{In order to satisfy the assumption about the independence between the training and test set in the cross-validation, we could not exceed 4-fold in two of the seven benchmark datasets, i.e H3K36me3\_TDH\_immune \& H3K36me3\_TDH\_other (see Supplementary Table 1).} cross-validation was performed on each of the seven datasets. Here, the results are shown by type of experiments (H3K36me3 \& H3K4me3). Formally, the accuracy can be written $1-\left(\sum_{m\in\text{ test set}}E_m\ /\ \sum_{m\in\text{ test set}} |H_m|\right)$.

\paragraph{Learning of tuning parameters.} In section \ref{supervised.learning} we described two methods for learning the tuning parameters of segmentation models. Based on results shown in supplementary, for the rest of this section, the parameters of the models compared on H3k36me3 datasets are learned through the \textit{constant $\lambda$} method. the parameters of the models compared on H3k4me3 datasets are them learned through the \textit{linear $\lambda$} method.
\paragraph{The over-dispersion exhibited by count data under a Poisson noise model can be effectively reduced using a negative binomial or a Gaussian transformed noise model.} \begin{figure}[ht]
\centerline{\includegraphics[scale=0.55]{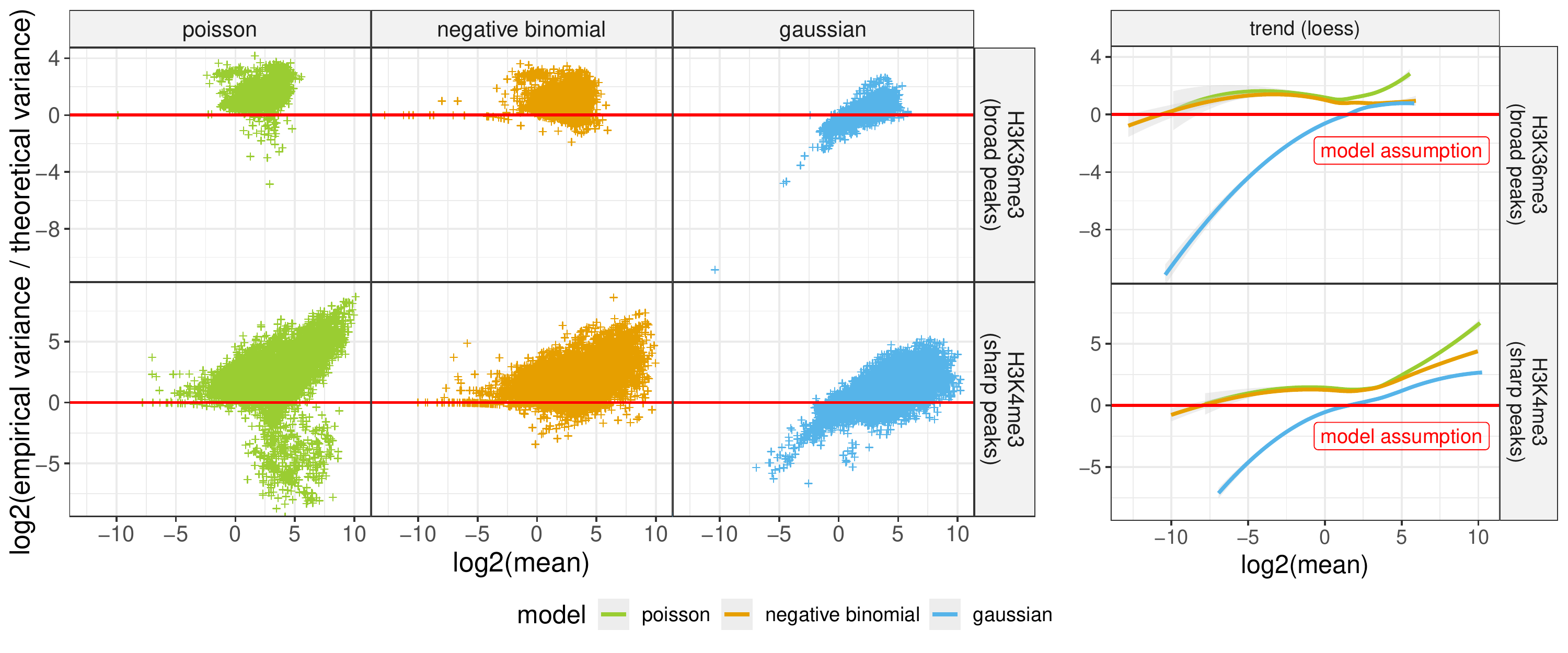}}
\caption{The over-dispersion exhibited by count data under a Poisson noise model (\crules[poisson]{0.25cm}{0.25cm}) can be effectively reduced using a negative binomial (\crules[negbin]{0.25cm}{0.25cm}) or a Gaussian transformed (\crules[gaussian]{0.25cm}{0.25cm}) noise model. The red indicator line (\crules[red]{0.25cm}{0.25cm}) stands for the equality of the theoretical and empirical variances. (\textbf{Left}) Observations above this line stand for over-dispersed count data. Observations under this line stand for under-dispersed count data. (\textbf{Right}) A loess curve has been computed on the observations for each noise model.}
\label{alieh:empirical.variance.vs.theoretical.variance}
\end{figure} Initially, we wanted to validate the presence of over-dispersion in count data following a Poisson distribution. In a second step, we wanted to confirm that alternative noise models such as the negative binomial or the Gaussian one, following an Anscombe transformation of the counts for the latter, could allow us to reduce this over-dispersion. A simple way to highlight the over-dispersion is to plot the $\log_2\text{-ratio}$ of the empirical and theoretical variances of count data against their mean. If the $\log_2\text{-ratio}$ is positive, the distribution of count data exhibits over-dispersion. If it is negative, the distribution of count data exhibits under-dispersion. If it is null, the dispersion of the count data does not show inconsistency with respect to the noise model. In figure \ref{alieh:empirical.variance.vs.theoretical.variance}, each observation corresponds to a segment from the segmentations selected during the cross-validation procedure for the 2752 coverage profiles. The segmentation were computed using \textit{CROCS} \includegraphics[scale=0.05]{figure_crocs.png} with \textit{gfpop} $+$ the PDPA model as solver. For each of the selected segments we estimated the empirical and theoretical variances. In the case of the Poisson noise model, the estimated theoretical variance is formally written $\hat{\sigma}^2 = \hat{\mu}$, where $\hat{\mu}$ stands for the estimation of the mean of count data in this segment. For the negative binomial one it is formally written $\hat{\sigma}^2 = \hat{\mu}+\phi^{-1}\hat{\mu}^2$, where $\phi$ stands for the dispersion parameter learned during the cross-validation procedure. For the Gaussian one, the theoretical variance is assumed constant across the segments. We estimated it using the mean squared error computed over all segments. In figure \ref{alieh:empirical.variance.vs.theoretical.variance} we can see that most of the count data under the Poisson noise model exhibit over-dispersion in both H3k36me3 and H3K4me3. Indeed, when looking at the segments with on average more than one count per base $(\log_2(\text{mean})>0)$, the loess curve computed on them is always above the red line which stands for the equality of the two variances. It is also in constant growth which suggests that the over-dispersion becomes larger as the mean of segments increases. To a lesser extent, count data under Poisson noise model with a large mean exhibit also under-dispersion in H3K4me4 datasets. In both H3K36me3 and H3K4me3 datasets, the loess curve of the Gaussian transformed noise model is always below the one from Poisson noise model while the loess curve of the negative binomial noise model is either indistinguishable or below the one of the Poisson noise model. Therefore, the use of these two alternatives noise models helps to partially correct this over-dispersion phenomenon although the reduction seems to be greater with the Gaussian transformed noise model on H3K4me3 datasets. We also note that a small part of segments with count data under Gaussian transformed noise exhibit under-dispersion. From a modeling perspective these count data are a part of the background noise, regions where a lot of genomic positions have a null associated count.
\begin{figure}
\centerline{\includegraphics[scale=0.6]{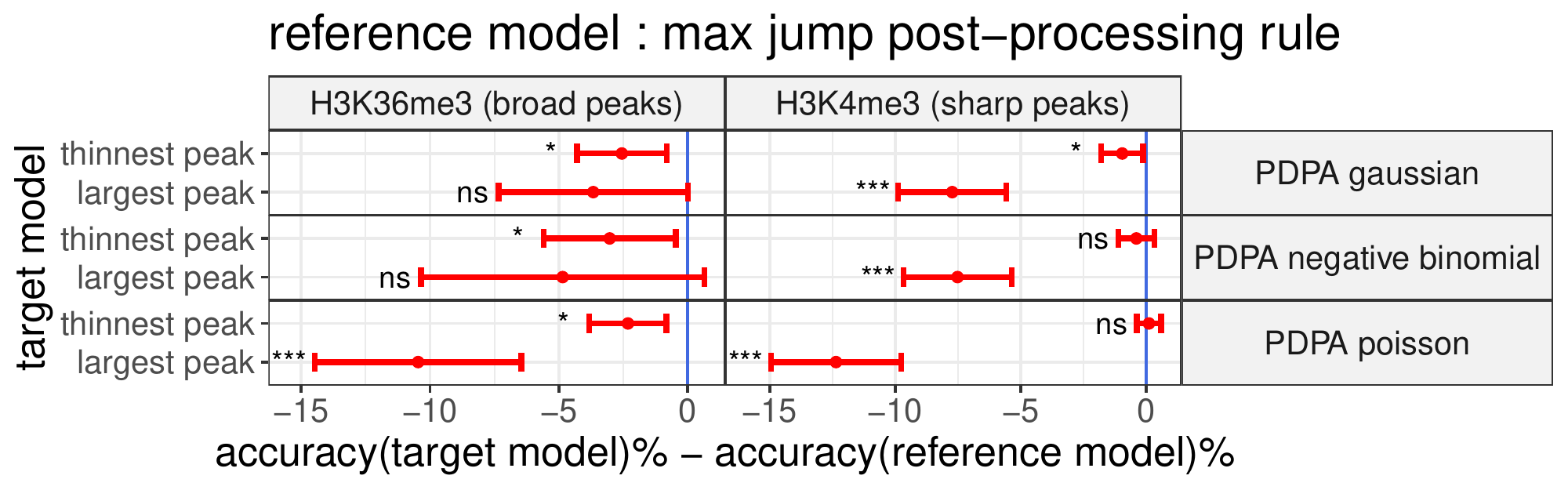}}
\caption{Max jump is the more accurate rule for inferring the peaks in segmentations obtained through a PDPA model. The difference in mean accuracy and its 95\% CI computed on the test folds pooled by types of experiment (H3K36me3 \& H3K4me3) are shown in red (\crules[red]{0.25cm}{0.25cm}). If the difference in mean accuracy is negative (left side of the blue indicator line \crules[blue]{0.25cm}{0.25cm}), the max jump rule is better in average than the target rule. The statistical significance of the difference in mean accuracy (paired t-test) is summarized in the following way: non significant (ns) means adjusted p-value $>0.05$; * means adjusted p-value $\leq0.05$; ** means adjusted p-value $\leq0.01$; *** means adjusted p-value $\leq0.001$.} 
\label{alieh:res.rules}
\end{figure}
\paragraph{Max jump is the more accurate rule for inferring the peaks in segmentations obtained through the PDPA model.} \label{res.maxjump} We wanted to compare the peak detection accuracy of the new rule we propose (\textit{max jump}) to select the changepoints corresponding to the start and end of the peaks with others (\textit{largest peak} \& \textit{thinnest peak}) which have an equivalence in \citet{toby20}. In the user guide of how to create labels in ChIP-seq coverage profiles \citep{toby16}, the authors strongly advise to label peaks which are obviously up with respect to the background noise. Hence, we expected that the \textit{max jump} rule, which sets the start and end of the peaks on the change with the largest mean-difference, performs at least as well as the other two rules. In figure \ref{alieh:res.rules}, we look at the difference in mean accuracy between each model with either the \textit{largest peak} or \textit{thinnest peak} rule, denoted target models, against the same model with the \textit{max jump} rule, denoted reference model. In agreement with our expectation, we observe that for the different models in both H3K36me3 \& H3K4me3 datasets, the mean accuracy of the \textit{max jump} rule is greater than the mean accuracy of the \textit{largest peak} rule (4.85\% to 12.36\% more accurate on average). Except for the PDPA model with a Poisson noise in H3K4me3 (0.11\% less accurate on average), the mean accuracy of the \textit{max jump} rule is also greater than the mean accuracy of the \textit{thinnest peak} ( 0.38\% to 3.03\% more accurate on average). In order to test the statistical significance of the difference in mean accuracy observed between the target and the reference model we performed a paired t-test. The accuracy of each fold were previously pooled by type of experiments as it is suggested in figure \ref{alieh:res.rules}. After correcting the p-values of the paired t-test with the Benjamini \& Hochberg method, 8 differences in mean accuracy were still significant (adjusted p-value < 0.05). As a result of these observations, for the next comparison we will infer the peaks in the segmentations obtained through the PDPA model using the new \textit{max jump} rule we propose.
\paragraph{PDPA model with a negative binomial or a Gaussian transformed noise is more accurate than previous state-of-the-art}\label{res.heuristic.vs.updown.vs.pdpa} \begin{figure}
\centerline{\includegraphics[scale=0.6]{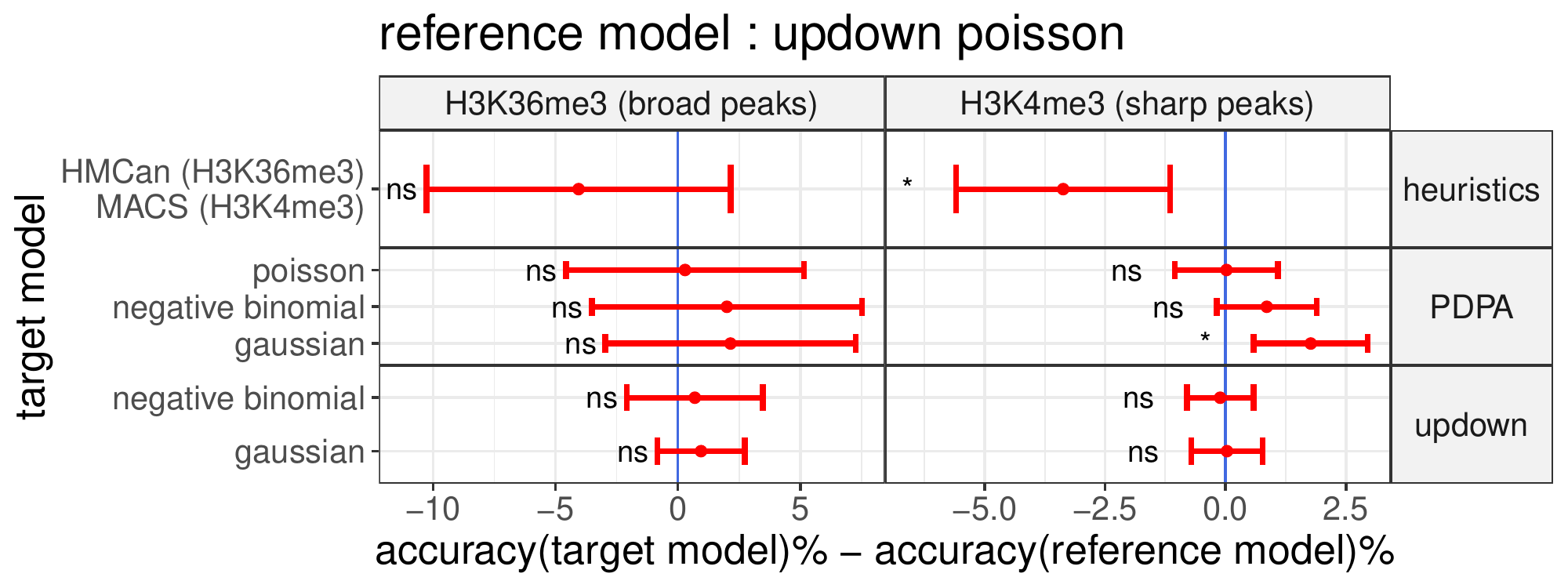}}
\caption{PDPA model with a negative binomial or a Gaussian transformed noise is more accurate than previous state-of-the-art. The difference in mean accuracy and its 95\% CI computed on the test folds pooled by types of experiment (H3K36me3 \& H3K4me3) are shown in red (\crules[red]{0.25cm}{0.25cm}). If the difference in mean accuracy is negative (left side of the blue indicator line \crules[blue]{0.25cm}{0.25cm}), the Up-Down model with a Poisson noise is better in average than the target model. The statistical significance of the difference in mean accuracy (paired t-test) is summarized in the following way: non significant (ns) means adjusted p-value $>0.05$; * means adjusted p-value $\leq0.05$; ** means adjusted p-value $\leq0.01$; *** means adjusted p-value $\leq0.001$.}
\label{alieh:res.comp}
\end{figure} 
We wanted to compare the peak detection accuracy of the Up-Down model with a Poisson noise \footnote{model built on natural assumptions to detect peaks in count data and actual state-of-the-art on H3k36me3 and H3k4me3 datasets} against other segmentation models such as the PDPA or Up-Down model with either a negative binomial or a Gaussian transformed noise. The heuristics HMCan and MACS have already been compared to the Up-Down model with a Poisson noise in \citet{toby20}. We included them again as a baseline from the bioinformatics literature. Both of them use a threshold that affects their peak detection accuracy and whose learning is also described in the previous cited study. Because we saw in previous results that a negative binomial or Gaussian transformed noise effectively reduces the over-dispersion exhibited by count data under a Poisson noise, we expected that the PDPA or Up-Down model with these alternative noises will improve the peak detection accuracy on the test set. In figure \ref{alieh:res.comp} we look at the difference in mean accuracy between the Up-Down model with a Poisson noise, denoted reference model, against other segmentation models and heuristics, denoted target models. In agreement with our expectation, we can see that the PDPA model with a negative binomial noise has a mean accuracy greater than the reference model in both H3K36me3 and H3K4me3 datasets (respectively 2.0\% and 0.86\% more accurate). It has also a greater mean accuracy with a Gaussian transformed noise (respectively 2.15\% and 1.77\% more accurate). As described previously, we performed a paired t-test to test the statistical significance of the difference in mean accuracy between each target and the reference model. After correcting the p-values of the paired t-test, the PDPA model with a Gaussian transformed noise was still significant (adjusted p-value < 0.05). Note that the PDPA model with a Poisson noise has a mean accuracy similar to reference model (difference in mean accuracy $<$ 0.5\% in both datasets). Thus, the improvement in accuracy cannot be attributed solely to the PDPA model $+$ the \textit{max jump} rule but also to the distribution chosen for the noise. In disagreement with our expectation, with the Up-Down model the use of alternative noise distributions does not improve the accuracy compared to the Poisson one (differences in mean accuracy $<$ 1\% in H3K36me3 and $<$ 0.1\% H3K4me3).  

\section{Conclusion and perspectives}

\paragraph{Supervised algorithm for learning segmentation models.} We developed the \textit{CROCS} \includegraphics[scale=0.05]{figure_crocs.png} algorithm that computes, for any segmentation model with constant penalty $\lambda$ along the changepoints, all optimal segmentations between two peak bounds. This set of segmentations is essential to compute the error rate function, which is in turn used in the supervised approach for learning the tuning parameters of the segmentation models (see Section \ref{supervised.learning}).

\paragraph{Modeling of over-dispersed ChIP-Seq count data.} We saw in figure \ref{alieh:empirical.variance.vs.theoretical.variance} that count data under a Poisson noise model exhibit overdispersion in H3k36me3 and H3K4me3 datasets. We showed that this overdispersion can be effectively reduced in these datasets using either a negative binomial or a Gaussian transformed noise model.

The use of a negative binomial noise model implies that we must be able to estimate a suitable value for the $\phi$ dispersion parameter. In section \ref{supervised.learning} we proposed to learn it jointly with the penalty of the segmentation model directly on the labeled coverage profiles. More precisely, a constant $\phi$ is selected because it minimizes the label errors of the training set. The negative binomial combined with the constant dispersion parameter allows the phenomenon of over-dispersion to be slightly reduced. 

With the Gaussian noise model there are no additional parameters than the penalty of the segmentation model to set. This is an advantage compared to the negative binomial one. In this study, in order to satisfy the Gaussian proprieties, we transformed the count data with the Anscombe transformation which is highly appreciated for its variance stabilization properties. Gaussian transformed noise model allowed to reduce the over-dispersion even more efficiently than the negative binomial noise model on the H3K4me3 dataset and as efficiently on the H3K36me3 datasets, while being simpler to implement.

\paragraph{Segmentation models for peak detection in ChIP-Seq count data.} In figure \ref{alieh:res.comp} we saw that when combining the PDPA model with a negative binomial or a Gaussian transformed noise it is possible to improve upon the natural and current state-of-the-art on the peak detection accuracy, the Up-Down model with a Poisson noise, in both H3K36me and H3K4me3 datasets. We argue that this improvement is likely explained by the ability of the negative binomial and Gaussian distribution to model the over-dispersion as illustrated in figure \ref{alieh:empirical.variance.vs.theoretical.variance}. In summary, we believe that the better we model dispersion the better we improve the accuracy of the segmentation model.

The PDPA model seems to capture more subtle changes in count data than the Up-Down one which have sometimes a poor fit to the signal (see Figure \ref{alieh:presentation.models} \& \ref{alieh:PDPA.rules.explain.fig}). One major issue of the PDPA model is its output segmentation which doesn't have a straightforward interpretation in terms of peaks compared to the Up-Down one. The introduction of the \textit{max jump} rule (see Section \ref{alieh:PDPA.rules.explain}), which have shown to perform at least as well as rules proposed in \cite{toby20} (\textit{thinnest peak} \& \textit{largest peak}), helped to correct this weakness (see Figure \ref{alieh:res.rules}). 

Still in figure \ref{alieh:res.comp}, we also saw that the Up-Down model with a negative binomial or a Gaussian transformed noise, which reduce the over-dispersion phenomenon, doesn't improve the accuracy upon the Up-Down model with a Poisson noise. One hypothesis to explain these results is that the constraints, which lead to the reduction of the space of optimal reachable segmentations with the Up-Down model, also reduce the probability of adding biologically uninformative changepoints induced by the over-dispersion. Consequently, the Up-Down model has the advantage to be a model with good internal over-dipsersion resistance properties but is bounded by its poor adaptability to the signal. 

\paragraph{Future work.} Figure 5 suggests that with both negative binomial and Gaussian transformed noise model the over-dispersion could be further reduced. Regarding the negative binomial noise model, one could think about predicting a specific-dispersion parameter for each coverage profile. Furthermore, the literature about Gaussian transformations is wide and a comparative study integrating segmentation models with different transformations for count data, e.g. the Box–Cox transformation or arcsine square root transformation, would also be an interesting avenue for future work. 

In this study we explored two different segmentation models, the unconstrained segmentation model (PDPA) and a constrained segmentation model where each non-decreasing change is followed by an non-increasing change (Up-Down). The \textit{gfpop} method makes it possible to model changepoints even more precisely by constraining for example the minimum size of jumps or the minimum size of segments. It would be interesting in future work to test other constrained models or include auto-correlation \citep{romano20, cho15} in the context of the peak detection problem for ChIP-seq count data. 

\section{Reproducible Research Statement} We have created a dedicated \textit{GitHub} repository with the code and data necessary to reproduce our figures and empirical results: \url{https://github.com/aLiehrmann/chip_seq_segmentation_paper/}
\bibliographystyle{abbrvnat}
\bibliography{references}
\includepdf[pages=-]{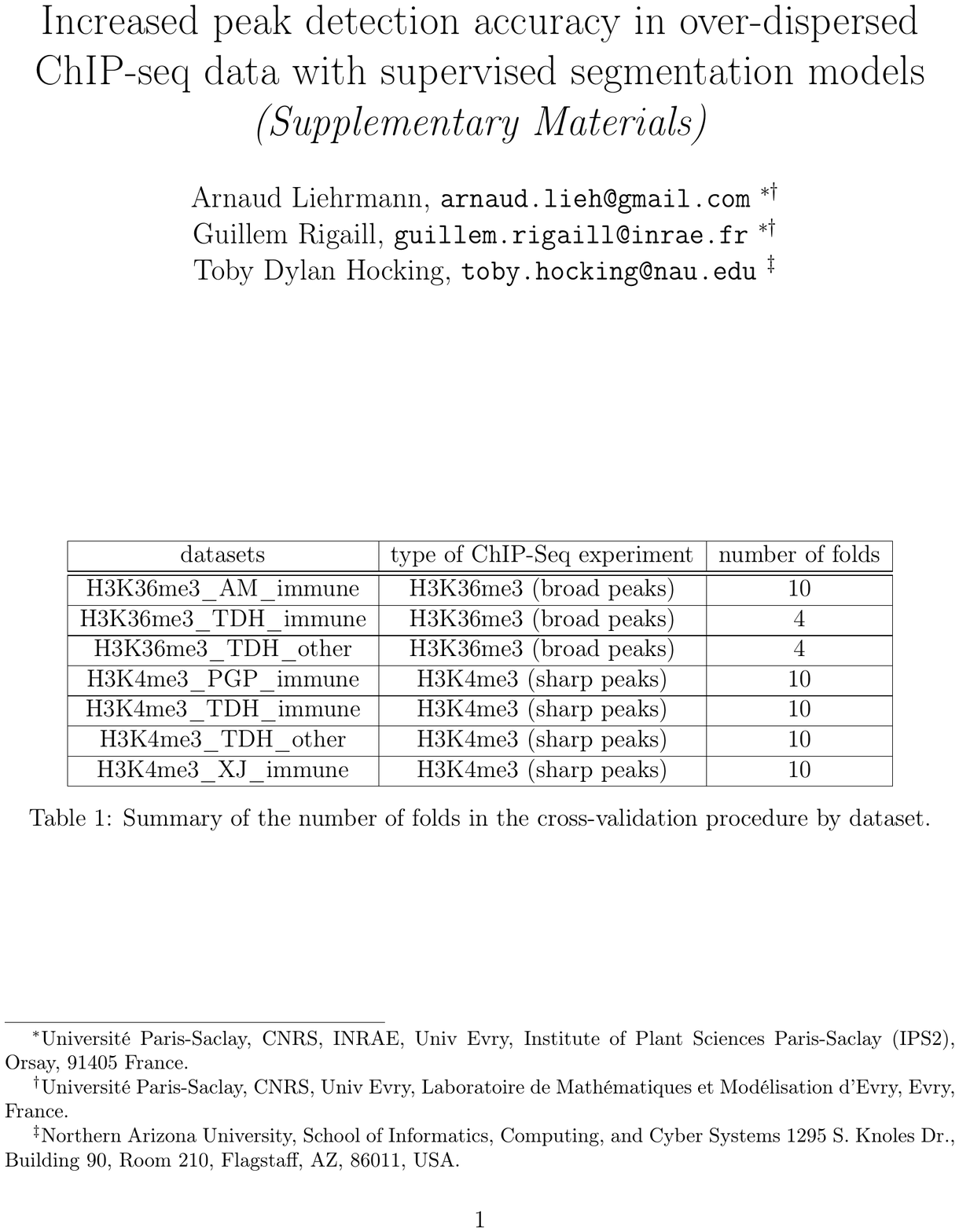}
\end{document}